\documentclass[12pt]{article}

\usepackage{scicite}
\usepackage{times}

\usepackage[utf8]{inputenc}
\usepackage[hidelinks]{hyperref}
\usepackage[]{graphicx}
\usepackage{amsmath,amssymb,xfrac,multirow}
\usepackage{float}
\usepackage{xcolor}
\usepackage{siunitx}
\usepackage{appendix}
\usepackage{miller}
\usepackage{scalerel}
\usepackage{verbatim}
\usepackage{cleveref}
\usepackage{soul}
\usepackage[T1]{fontenc}

\usepackage{caption}
\usepackage{xparse}
\usepackage{microtype}

\Crefname{figure}{}{}

\renewcommand{\thetable}{\arabic{table}} 
\renewcommand{\thefigure}{\arabic{figure}}
\newcommand{\Fig}[2]{Fig.~\ref{#1}\IfValueT{#2}{\MakeUppercase{#2}}}
\newcommand{\Figs}[2]{Figs.~\Cref{#1}\IfValueT{#2}{\MakeUppercase{#2}}}
\newcommand{\Figure}[2]{Figure~\ref{#1}\IfValueT{#2}{\MakeUppercase{#2}}}
\newcommand{\Figures}[2]{Figures~\Cref{#1}\IfValueT{#2}{\MakeUppercase{#2}}}
\newcommand{\fig}[2]{fig.~\ref{#1}\IfValueT{#2}{\MakeUppercase{#2}}}
\newcommand{\figs}[2]{figs.~\Cref{#1}\IfValueT{#2}{\MakeUppercase{#2}}}
\newcommand{\Tab}[1]{table~\ref{#1}}

\usepackage{nameref}
\newenvironment{sciabstract}{%
\begin{quote} \bf}
{\end{quote}}

\title{\centering Secondary Grain Boundary Dislocations Alter Segregation Energy Spectra}

\author{
Xinren Chen$^{1}$, William Gonçalves$^2$, Yi Hu$^{1}$, Yipeng Gao$^{1}$,\\ 
Patrick Harrison$^3$, Saurabh Mohan Das$^{1}$, Gerhard Dehm$^1$, Baptiste Gault$^{1,4}$,\\  
Wolfgang Ludwig$^2$, Edgar Rauch$^3$,  Xuyang Zhou$^{1\ast}$, Dierk Raabe$^1$\\
\\
\\
\normalsize{$^{1}$Max-Planck-Institut for Sustainable Materials, D\"usseldorf 40237, Germany}\\
\normalsize{$^{2}$Université Lyon I, MATEIS, INSA Lyon, CNRS UMR 5510, Villeurbanne 69621, France}\\
\normalsize{$^{3}$Université Grenoble Alpes, CNRS, Grenoble INP, SIMaP, Grenoble F-38000, France}\\
\normalsize{$^{4}$Department of Materials, Royal School of Mines, Imperial College, London, UK}\\
\normalsize{$^\ast$To whom correspondence should be addressed; E-mail:  x.zhou@mpie.de}
}

\date{}

\begin{document} 

\baselineskip24pt
\maketitle 

\begin{sciabstract}

Grain boundaries (GBs) trigger structure-specific chemical segregation of solute atoms. 
According to the three-dimensional (3D) topology of grains, GBs — although defined as planar defects — cannot be free of curvature.
This implies formation of topologically-necessary arrays of secondary GB dislocations.
We report here that these pattern-forming secondary GB dislocations can have an additional and, in some cases, even a much stronger effect on GB segregation than defect-free GBs. 
Using nanoscale correlative tomography combining crystallography and chemical analysis, we quantified the relationship between secondary GB dislocations and their segregation energy spectra for a model Fe-W alloy. 
This discovery unlocks new design opportunities for advanced materials, leveraging the additional degrees of freedom provided by topologically-necessary secondary GB dislocations to modulate segregation.

\end{sciabstract}  

% \begin{refsegment}
Metals and alloys usually consist of crystalline grains that fill three-dimensional (3D) space. 
These grains meet at junctions to form piecewise planar defects known as grain boundaries (GBs).
Due to topological reasons, GBs of real polycrystalline materials exhibit curvature \cite{Chen2024}, to reconcile force equilibrium at nodes and junctions with pore-free, space-filling topology of the non-platonic shaped grains \cite{Ashby1970}. 
These curvature variations are realized by the build-up of disconnections, including steps and secondary GB dislocations \cite{Bishop1968,Hirth1973,Rajabzadeh2014,Meiners2020a}.
Secondary GB dislocations, which are GB line defects with dislocation character and often form patterns with regular spacing (see \Fig{fig:1}{A-C}), are well known to play particularly important roles in GB migration \cite{Qiu2024,He2023}, sliding \cite{Wang2022}, and rotation \cite{Tian2024}. 

Here we show that secondary GB dislocations also alter GB segregation. 
Impurities or solutes tend to segregate to GBs to minimize their total free energy, as originally described by the Gibbs adsorption isotherm \cite{Mclean1957,Luo2011,Nie2013,Yu2017}. 
Due to their localized strain fields, the pattern-forming secondary GB dislocations can induce additional — and sometimes even much stronger — GB segregation compared to the segregation in defect-free GB structures. 
More specific in the model system FeW studied here we find that this effect can result in a twofold increase in the composition of GB segregation compared to a defect-free GB segment. 
Interactions between secondary GB dislocations and solute atoms significantly alter the mechanical and physical properties of materials \cite{Zhu2020}. 
Capturing these complex solute–GB interactions at highest resolution scale is critical for understanding GB decoration and unlocking atomic-scale engineering of alloys \cite{Raabe2014}.

There are four major challenges for the experimental characterization of solute-GB interactions. 
Firstly, GBs often deviate from flat planes and exhibit undulating 3D structures \cite{Zhang2017,Xu2021}. 
Real GB structures invariably break down into facets and regular patterns formed by secondary GB dislocations, to accommodate topological constraints \cite{Ashby1970,Chen2024}. 
These curved surfaces contain defects that significantly alter solute segregation behavior and create local solute enrichment patterns \cite{Hirth1973,Yao2013,Liebscher2018,Zhao2020,Priedeman2020,Zhou2021,Zhou2023}. 
To capture the full structural and chemical characteristics of a GB, one must conduct these measurements in 3D. 
Accurate quantification of solute segregation requires atomic-scale spatial resolution and high chemical sensitivity, as solute segregation is confined, typically to only a few atomic layers from the GB interface \cite{Luo2011, Nie2013, Yu2017, Yang2020, Tehranchi2024, Langenohl2023}.
In addition, a complete representation for any GB requires the measurement of five kinematic degrees of freedom \cite{Harrison2024}: three to describe the misorientation between adjacent grains and two to define the orientation of the GB segment plane. 
Finally, different types of GBs exhibit different segregation behaviors, sometimes varying by more than an order of magnitude. 
To conduct a statistically relevant analysis, it is thus essential to have the capability to study multiple GBs with a range of misorientations and GB plane orientations, if possible, simultaneously for high throughput.

Here, we apply four-dimensional scanning transmission electron microscopy (4DSTEM) tomography \cite{Harrison2022} to gain structural insights and use atom probe tomography (APT) for compositional analysis on a model body-centered cubic (BCC) Fe-1 at.\% W alloy.
This correlative nanoscale tomography approach enables the measurement of five kinematic degrees of freedom for any GB \cite{Harrison2024} and quantifies near atomic-scale solute-GB interactions at defect-rich GBs.
4DSTEM tomography efficiently detects defect contrasts at GBs, eliminating the need for meticulous crystal tilting required by conventional transmission electron microscopy (TEM) and simplifying the characterization of randomly oriented GBs with undulating 3D structures.
By thoroughly examining both the structural and chemical characteristics across 12 Fe GBs, we provide quantitative analyses for revealing the correlations between GB structure and W segregation. 
This knowledge is essential to quantify the interactions between secondary GB dislocations—necessitated by curvature variations—and alloying elements, which opens design opportunities for engineering future materials.

\subsubsection*{Correlative tomography for three-dimensional crystallography and chemistry} \label{Correlative}

\begin{figure}[H]
\centering
\includegraphics[width=16cm]{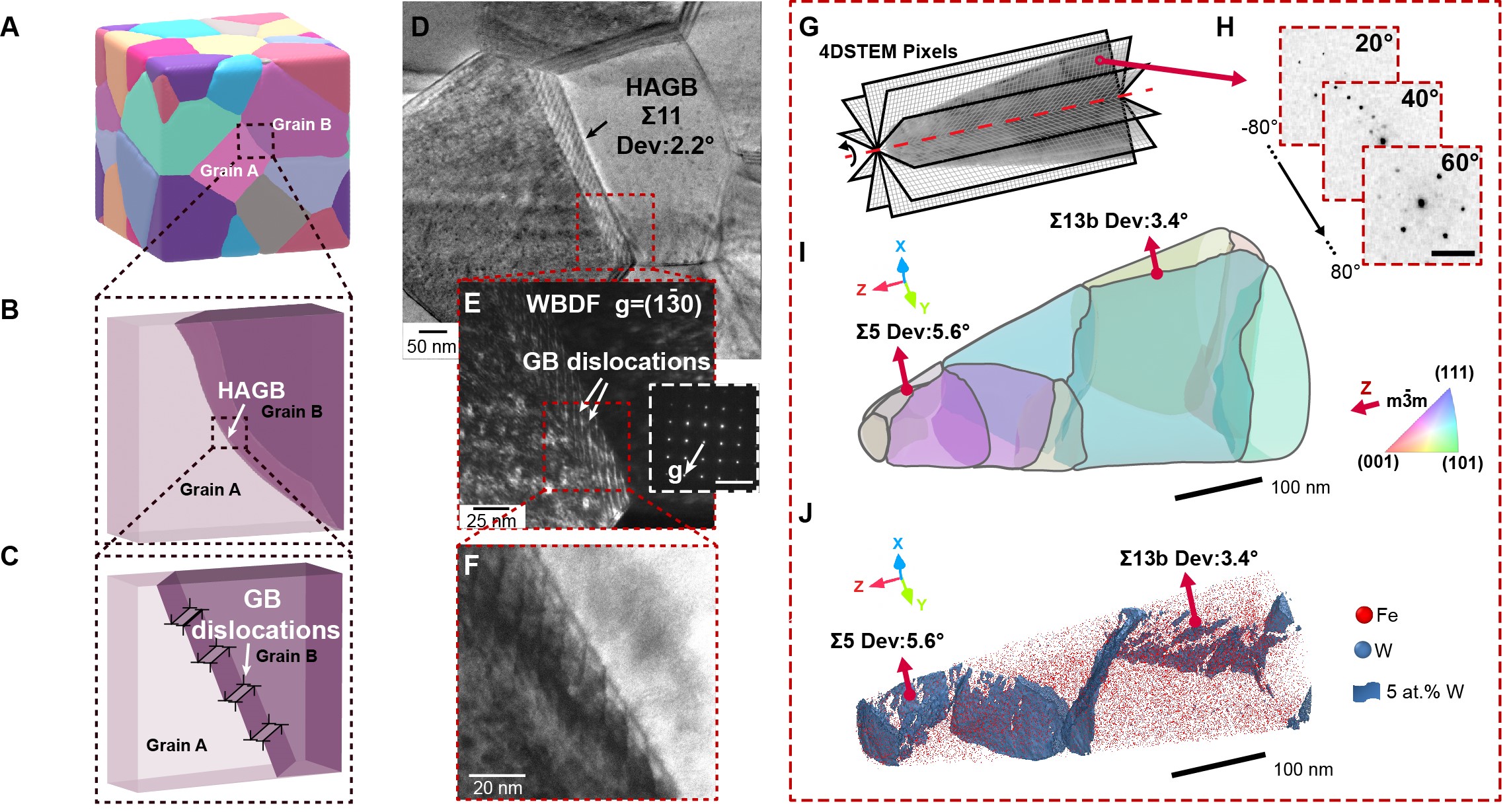}
\caption{\textbf{Correlative tomography characterization of the Fe-1 at.\% W specimen.}
    (\textbf{A}) Illustration of grain boundaries (GBs) in a three-dimensional (3D) polycrystalline structure.
    (\textbf{B}) Magnification of a selected high-angle GB (HAGB) showing curved GB plane.
    (\textbf{C}) Illustration of steps and secondary GB dislocations on the GB.
    (\textbf{D}) Bright-field image of a transmission electron microscopy (TEM) lamella prepared from the body-centered cubic (BCC) Fe-1 at.\% W specimen, featuring multiple GBs, one of which is labeled as a HAGB.
    This HAGB is a $\Sigma 11$ GB that deviates by \SI{2.2}{\degree} from the theoretical misorientation.
    (\textbf{E}) Corresponding weak-beam dark-field image of the same area, imaged under the two-beam condition with $\mathbf{g} = [1\bar{3}0]$. 
    The inset image in the right side of (E) displays the diffraction pattern collected from the left grain.
    Scale bar: 1/0.1 nm.
    (\textbf{F}) High magnification bright-field image displaying the step feature of the GB.
    (\textbf{G}) Illustration of four-dimensional scanning transmission electron microscopy (4DSTEM) tomography. 
    Each pixel in the 4DSTEM data incorporates a local nanobeam diffraction pattern.
    (\textbf{H}) Representative nanobeam diffraction patterns from the 4DSTEM datasets at different tilt angles for the grain indicated by red circle in (G). 
    Scale bar: 1/0.083 nm.
    (\textbf{I}) 3D crystallographic reconstruction of the grains in the Fe-W needle-shaped specimen, characterized via 4DSTEM tomography \cite{Harrison2022,Harrison2024}.
    Grains are colored based on the Euler angle representation of their crystallographic orientations relative to the Z-axis, serving as the tilting axis in TEM.
    The X-axis represents the thin film growth direction in the same coordinate system.
    (\textbf{J}) 3D chemical reconstruction of the same specimen shown in (I) by atom probe tomography (APT), illustrating the spatial distribution of Fe (red) and W (cornflower blue) atoms with a superimposed 5 at.\% W isosurface.
    The 5 at.\% isosurface represents the region (voxels) containing 5 or more at.\% W.
    For a clear visualization, we aligned the APT reconstruction in the same perspective as the 4DSTEM tomography reconstruction shown in (I).
    }
    \label{fig:1}
\end{figure}

We prepared the model BCC Fe-1 at.\% W alloy thin film using physical vapor deposition (PVD) accompanied by a heat treatment at \SI{500}{\degree C} for \SI{240}{minutes} to activate diffusion and facilitate solute segregation at GBs and the secondary defects on them.
The average grain size is approximately \SI{134}{nm}, with equiaxed grains and a weak \hkl{111} fiber texture in the growth direction (see~\fig{fig:MethodPVD}{}).
Periodic intensity contrasts in the weak-beam dark-field TEM images of high-angle GBs (HAGBs) typically stem from secondary GB dislocations, particularly in those nearly matching the coincidence site lattice (CSL) relationship \cite{Grimmer1974}.
\Figure{fig:1}{D-F} present the bright-field and weak-beam dark-field images of a $\Sigma 11$ GB that has a \SI{2.2}{\degree} deviation from the theoretical misorientation.
We confirmed that secondary GB dislocations are the source of the periodic intensity contrast, observable under two-beam conditions with $\mathbf{g} = [1\bar{3}0]$ (see \Fig{fig:1}{E}).

To further elucidate the interactions between solutes and defects (specifically secondary GB dislocations), we employed our recently introduced nanoscale tomography approach \cite{Harrison2022,Harrison2024}.
\Figure{fig:1}{G-I} shows a 3D crystallographic reconstruction of 11 grains and 12 GBs in a needle-shaped Fe-W specimen, providing a meaningful reference for the GB segregation landscape in materials with defect-containing GBs. 
We colored each grain based on its Euler angle representation of the crystallographic orientation relative to the Z-axis (see \Tab{tab:GrainEuler}), along which the specimen is tilted for tomography (see \Fig{fig:1}{G}), with the X-axis aligned parallel to the thin film growth direction.
The orientation was determined by analysis of the nanobeam diffraction patterns that were systematically collected during the 4DSTEM scans. 
\Figure{fig:1}{H} displays three example diffraction results taken at different tilts.
Our method enables the mapping of local normal to the GB plane between adjacent grains, as demonstrated in \fig{fig:GBPlane}{}.
Details on the sample preparation and data processing are available in the Supplementary Materials Section \nameref{Sample}, Section \nameref{TEM}, \fig{fig:Method4DSTEMAPT}{A}, and video.S1.

We analyzed the composition of this same specimen by APT.
\Figure{fig:1}{J} displays the spatial distribution of Fe as individual red dots, and a set of isocomposition surfaces that delineates regions containing 5 or more at.\% W.
The W enrichment closely correlates with the location of GBs identified in the 3D crystallographic reconstruction in \Fig{fig:1}{I}, evidencing preferential W segregation to GBs.
We focus particularly on the following two GBs with a near-$\Sigma$ relationship that exhibit discontinuous W segregation patterns.

\subsubsection*{Quantitative analysis of secondary grain boundary dislocations and chemical segregation} \label{LinePattern}

To quantify segregation patterns, a 3D compositional mapping of W within the APT reconstruction was calculated with a resolution of $0.5 \times 0.5 \times 0.5 \, \text{nm}^3$ and a delocalization parameter of \SI{3}{nm} \cite{Hellman2003}.
The compositional map plotted in \Fig{fig:2}{A} shows that the segregation of W at GBs forms a discontinuous pattern rather than a uniform planar distribution.
We selected two GBs for detailed analysis: GB1 between grain $\alpha_1$ and grain $\alpha_2$, characterized as a near $\Sigma 5 \hkl[100]$ GB with a deviation of \SI{5.6}{\degree} from its ideal misorientation, and GB2 between grain $\alpha_4$ and grain $\alpha_5$, identified as a near $\Sigma 13b \hkl[111]$ GB with a \SI{3.4}{\degree} deviation, illustrated in \Fig{fig:2}{B-E} and \Fig{fig:2}{F-I}, respectively.

\Figure{fig:2}{B} presents two mappings of the local normal to the GB plane between adjacent grains for GB1: the bottom mapping corresponds to grain $\alpha_1$, and the top to grain $\alpha_2$.
The similarity in color representation across both mappings indicates that the orientations of the GB planes on both sides of the interface are similar, suggesting a near-symmetric tilt $\Sigma 5 \hkl[100]$ GB.
We identified secondary GB dislocations by constructing a virtual dark-field image based on the vector $\mathbf{g=[12\bar{1}]}$ for grain $\alpha_2$, as shown in \Fig{fig:2}{C}.
See more details in \fig{fig:MVDF}{}.
\Figure{fig:2}{D} displays a stairwell-like periodic segregation pattern at this GB, highlighted by a set of isosurfaces with a threshold of 2.5 at.\% W.
A one-dimensional (1D) composition profile with a bin size of \SI{1}{nm} was calculated along the cylinder positioned through the segregation pattern in \Fig{fig:2}{D}. 
From this profile, plotted in \Fig{fig:2}{E}, we quantify the level of segregation ranging from 1 to 4 at.\% W.
Note that this pattern exhibits a regular spacing of approximately \SI{22}{nm}, a spacing that closely matches the spatial distribution of secondary GB dislocations imaged by virtual dark-field in \Fig{fig:2}{C}. 
The maximum value in the local composition profile is twice that of the mean composition of the GB, which represents that of a defect-free GB segment. 
This direct correlation suggests a key role of secondary GB dislocations in altering solute segregation, in terms of mechanism, trapping depth, patterning, magnitude and kinetics.

\begin{figure}[H]
\centering
\includegraphics[width=16cm]{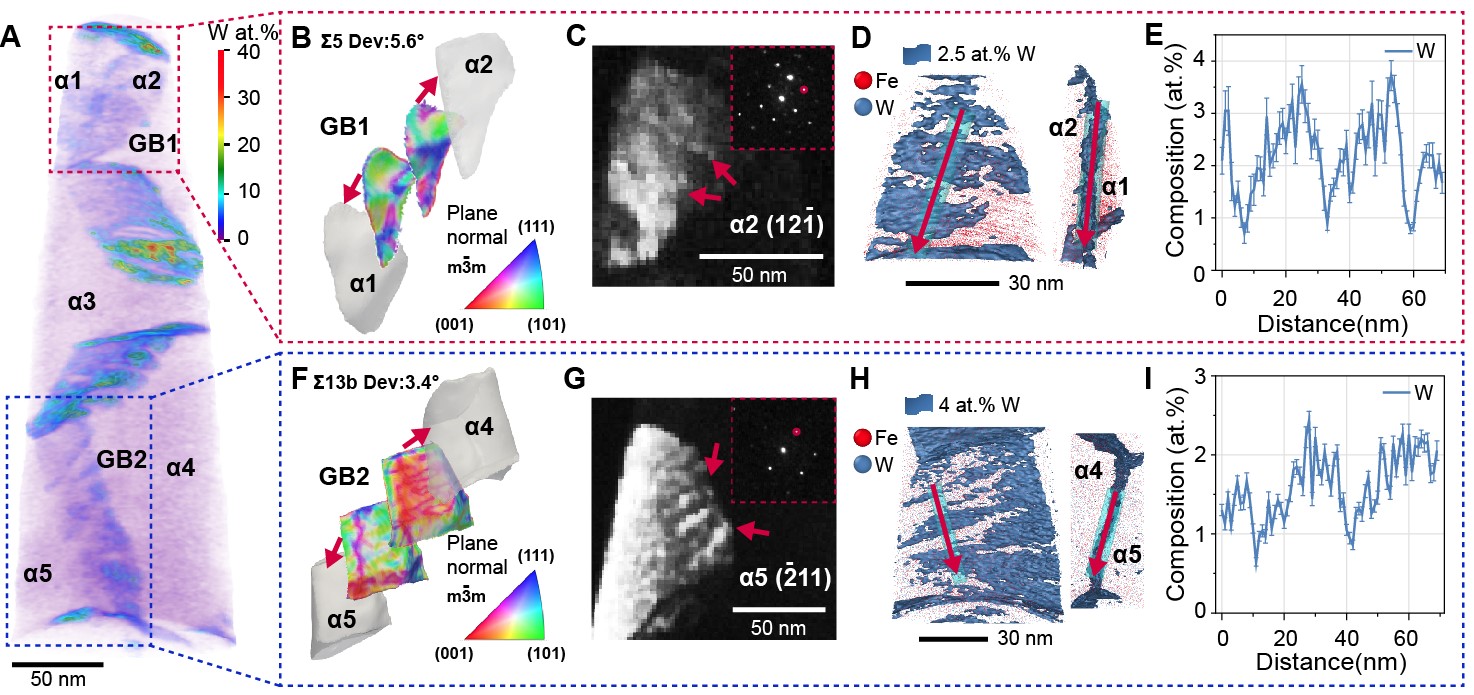}
    \caption{\textbf{Characterization of secondary GB dislocations and their linkage to segregation patterns in the Fe-1 at.\% W specimen.}
    (\textbf{A}) Compositional mapping of W at a resolution of $0.5 \times 0.5 \times 0.5 \, \text{nm}^3$ per voxel, enabling visualization of volume-specific variations.
    Parts of the grains and GBs have been labeled, with identities ranging from grain $\alpha_1$ to grain $\alpha_5$ and from GB1 to GB2 (see all labels in \fig{fig:GBPlane}{}).
    Correlative crystallographic and compositional quantitative analysis for two GBs (\textbf{GB1: B-E}) and (\textbf{GB2: F-I}). 
    Each set includes: \textbf{(B \& F)} orientation mappings of the local normal to the GB plane between adjacent grains (rendered in translucent grey), with two mappings provided for each GB, referenced to the respective grain involved; \textbf{(C \& G)} 4DSTEM virtual dark-field images; \textbf{(D \& H)} atom maps of Fe (colored red) and W (colored cornflower blue), with superimposed isosurfaces at 2.5 at.\% W and 4.0 at.\% W, respectively; \textbf{(E \& I)} W profiles along the red arrows shown in (E) \& (I).
    The insert images in (C) \& (G) display the nanobeam diffraction patterns with the vectors $\mathbf{g=[12\bar1]}$ for grain $\alpha_2$ and $\mathbf{g=[\bar211]}$ for grain $\alpha_5$ (highlighted by red circles), which were used to generate the virtual dark-
    }
    \label{fig:2}
\end{figure}
\begin{figure}[htbp]
\centering
    \ContinuedFloat % This indicates that this figure is a continuation
    \caption*{field images shown in (C, G).
    Two perspectives (\SI{90}{\degree} rotated clockwise from left to right) in (D \& H) are shown to illustrate the segregation patterns in 3D.    
    }
\end{figure}

We observe similar results at the $\Sigma 13b \hkl[111]$ GB, as shown in \Fig{fig:2}{F-I}.
Unlike the previous case, GB2 is an asymmetric tilt GB, with the plane normal for grain $\alpha_4$ aligning with $(\bar{1}\bar{1}2)$ and for grain $\alpha_5$ with $(\bar{1}10)$, as detailed in \Fig{fig:2}{F}.
Nevertheless, we still find a close match between the secondary GB dislocations shown in \Fig{fig:2}{G} and \fig{fig:VDFG5}{}, and the periodic segregation patterns (see \Fig{fig:2}{H}) and the composition profile plotted in \Fig{fig:2}{I}.
We also observed GB facets that can modulate solute segregation at GBs, similar to the reports in references \cite{Liebscher2018,Zhao2020,Zhou2023,Meiners2020b}. 
These were particularly noted at GB3 (see \fig{fig:GB3}{}).

\subsubsection*{Calculation of grain boundary segregation energy using the Langmuir-McLean isotherm} \label{McLean}

The relationship between solute segregation and segregation energy typically follows the Langmuir-McLean isotherm, as detailed in Equation \eqref{eq:1} \cite{Mclean1957,Langmuir1918,Wagih2020a,Murdoch2013},

\begin{equation}
\frac{X_{\text{GB}}}{1 - X_{\text{GB}}} = \frac{X_\text{B}}{1 - X_\text{B}} \exp \left( \frac{\Delta G_{\text{Seg}}}{RT} \right).
 \label{eq:1}
\end{equation}
Here, $\Delta G_{\text{Seg}}$ is the segregation energy of the solute atom, $R$ is the ideal gas constant, $T$ is the temperature, and $X_{\text{B}}$ and $X_{\text{GB}}$ are the compositions of solute in the bulk and at the GB, respectively. 
$X_{\text{GB}}$ can be further expressed as $\frac{\Gamma \cdot \Omega(1 - X_{\text{B}})}{t} + X_{\text{B}}$ \cite{trelewicz2009}, where $\Gamma$ represents the interfacial excess (IE), denoting the excess number of atoms per unit area at an interface, $t$ is the thickness of the GB, and $\Omega$ is the atomic volume of the solvent.

We mapped the IE values across all visible GBs in the APT dataset, with \Fig{fig:3}{A} providing examples of the integral profiles across GB1 used for these IE calculations, and \Fig{fig:3}{B} showing the mapped IE values \cite{Seidman1994, Zhou2022}.
We chose IE as it minimizes the influence of local magnification caused by different field evaporation behaviors arising from different compositions and grain orientations \cite{Miller1991,Zhou2022}.
Solute segregation at GBs typically spans only a few atomic layers, which is often smaller than the volume increment, potentially leading to inaccurate measurements of local segregation. 
Consequently, IE is employed for accurate quantification analysis.

\begin{figure}[H]
\centering
\includegraphics[width=16cm]{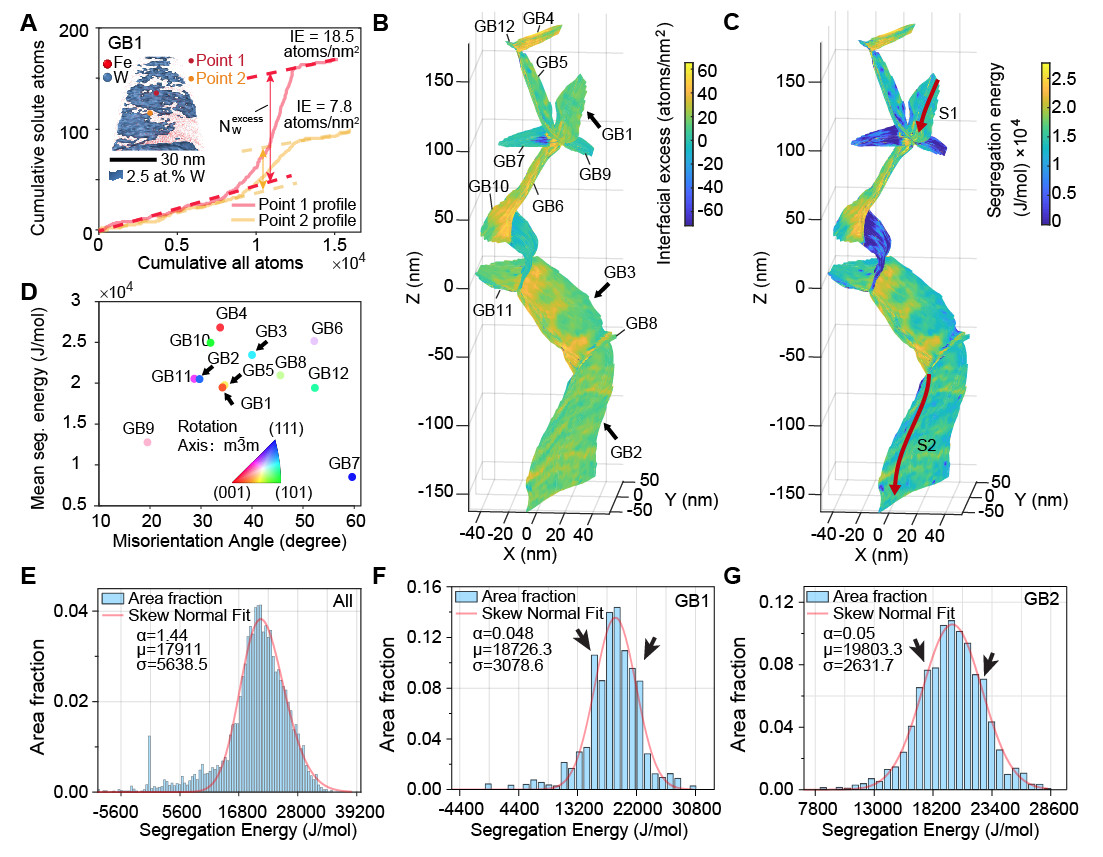}
    \caption{\textbf{Mapping and analysis of GB properties in the Fe-1 at.\% W specimen:}
    (\textbf{A}) Integral profiles across GB1 for quantifying the interfacial excess (IE) of W segregation at the Fe GB \cite{Seidman1994, Zhou2022}.
    We plot measurements from two points: Point 1 for high W regions (colored pink) and Point 2 for low W segregation regions (colored orange).
    In each plot, the solid lines show the cumulative relationship between all atoms and solute atoms, while the dashed lines are the fittings within the two grains adjacent to the GB plane.
    $N_{W}^{\text{excess}}$ represents the accumulation of excess atoms across the GB interface.
    IE values are calculated by dividing $N_{W}^{\text{excess}}$ by the corresponding interface areas, which are approximately $\SI{8.0}{\nm^2}$ for Point 1 and Point 2.
    We indicate the locations of Point 1 and Point 2 in the embedded image, consistent with \Fig{fig:2}{D}.
    (\textbf{B}) IE mapping on GBs, as identified from the correlative characterization in \Fig{fig:1}{I \& J}.
    We labeled the locations of all investigated GBs from GB1 to GB12 in (B).
    (\textbf{C}) Segregation energy mapping on the same GBs as in (B), calculated based on the Langmuir-McLean isotherm, Eq. \eqref{eq:1} \cite{Mclean1957,Langmuir1918}.
    }
    \label{fig:3}
\end{figure}
\begin{figure}[htbp]
\centering
    \ContinuedFloat % This indicates that this figure is a continuation
    \caption*{The locations where 1D line profiles of segregation energy values were measured are marked by red lines S1 and S2, details of which will be presented in the following section.  
    (\textbf{D}) Correlation between the misorientation angle and segregation energy of all investigated GBs: 
    This scatter plot displays the average segregation energy (in J/mol) for various GBs, plotted against their misorientation (in degrees). 
    Each point represents a different GB (GB1 to GB12), with colors indicating the rotation axis of misorientation related to the cubic crystal symmetry ($m\bar{3}m$), as shown in the color triangle legend.
    Experimental GB segregation energy spectra for (\textbf{E}) GB1, (\textbf{F}) GB2, and (\textbf{G}) all GBs.   
    The fitting plots, overlaid onto the histograms, generally follow the skew-normal function \cite{Wagih2020a} but with local deviations highlighted by the black arrows, see (E \& F). 
    The used fitting parameters are: characteristic energy ($\mu$) in kJ/mol, width ($\sigma$) in kJ/mol, and shape parameter ($\alpha$) along with each plot.  
    The sharp peak appearing near \SI{0}{J/mol} in panel (G) results from the contribution of the GBs without solute segregation and the inevitable minor inclusion of the bulk regions adjacent to the GBs.
    }
\end{figure}

\Figure{fig:3}{C} maps the segregation energy calculated from the IE, based on the Langmuir-McLean isotherm, Eq. \eqref{eq:1}.
While this isotherm does not account for solute-solute interactions, for more complex cases, models such as the Fowler-Guggenheim isotherm \cite{Fowler1939} or the Seah-Hondros isotherm \cite{Seah1973} could be similarly applied by adding an extra term to account for the interaction energy. 
Despite these complexities, we hypothesize that the general formulation suggested by the Langmuir-McLean isotherm (Eq. \eqref{eq:1}) remains applicable here, except for the case that it cannot account for any second order defect trapping on the GB, such as caused by GB dislocation structures.

We calculated the mean segregation energy for each GB. 
As plotted in \Fig{fig:3}{D}, it shows the same trends previously reported from low angles to \SI{60}{\degree} \cite{Herbig2014}, demonstrating a clear effect of misorientation on the segregation level.
Our measurements allow for much more refined analysis of the segregation energy landscape. 
We have generated segregation energy spectra, \Fig{fig:3}{E-G} for GB1, GB2, and all 12 GBs. 
The statistical distribution of solute segregation is generally close to a skew-normal function (see \Fig{fig:3}{E \& F} for GB1 and B2), as predicted by atomistic simulations \cite{Wagih2020a}, a feature that had not yet been confirmed before by experiment. 
The experimental segregation energy spectra of defect-containing GBs deviate from the ideal curve, as marked by black arrows in \Fig{fig:3}{E \& F}.

\subsubsection*{Estimation of the elastic energy contribution for grain boundaries containing secondary dislocations} \label{Stress}

\begin{figure}[H]
\centering
\includegraphics[width=16cm]{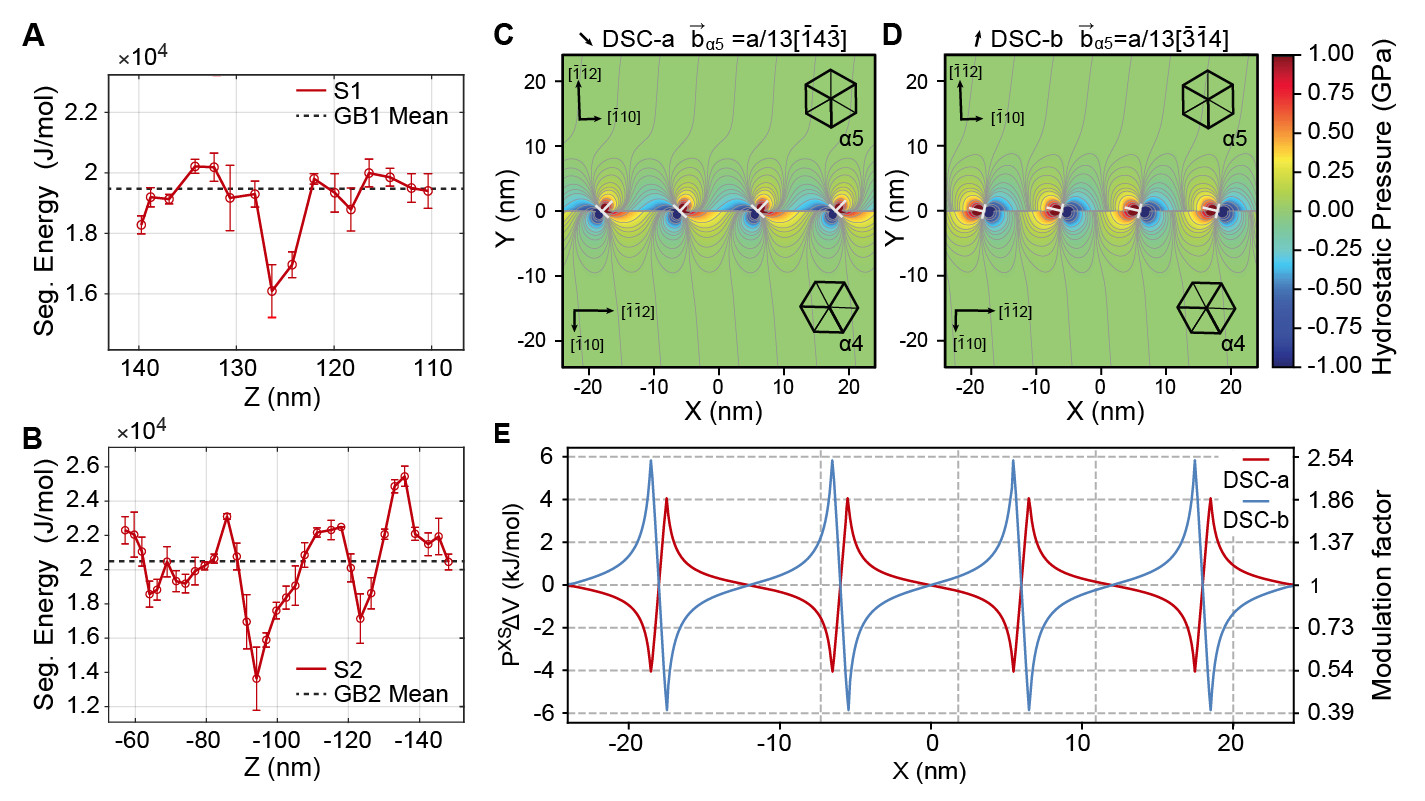}
    \caption{\textbf{Quantifying elastic energy for the $\Sigma 13b$ GB with dislocations using linear anisotropic elasticity theory.}
    (\textbf{A \& B}) Segregation energy profiles along the lines marked by red arrows in \Fig{fig:3}{C}: 
    \textbf{(A)} Line S1 at GB1 and \textbf{(B)} Line S2 at GB2, illustrating variations in segregation energy along these GBs.
    Horizontal dashed lines in (A) and (B) indicate the mean segregation energy of GB1 and GB2, respectively, serving as references.
    (\textbf{C \& D}) The stress field surrounds secondary GB dislocations for GB2 ($\Sigma 13b$, see \Fig{fig:2}{F-I}) with the displacement shift complete (DSC)-lattice vectors as Burgers vectors: (\textbf{C}) DSC-a: $\mathbf{b_{\alpha5}} = \frac{a}{13} [\bar14\bar3]$ and (\textbf{D}) DSC-b: $\mathbf{b_{\alpha5}} = \frac{a}{13} [\bar3\bar14]$.
    Here, $a$ is the lattice constant of BCC Fe.
    (\textbf{E}) The change in segregation energy ($P^\text{XS}\Delta V$) aligns along the GB, corresponding to (C \& D).
    Here, DSC-a and DSC-b have the same Burgers vectors as those in (C \& D).
    The modulation factor is defined as $\exp \left( \frac{-P^{\text{XS}} \Delta V}{RT} \right)$.
    }
\label{fig:4}
\end{figure}

To identify the impact of secondary GB dislocations on solute segregation, we analyzed the periodic variations in segregation energy along GB1 and GB2 (see \Fig{fig:3}{C}, Lines S1 and S2), using 1D composition profiles plotted in \Fig{fig:4}{A \& B}.
The gradient in these profiles can be attributed to the secondary GB dislocations visible in \Fig{fig:2}{C \& G}. 
The magnitude of the segregation energy change caused by these secondary GB dislocations can reach up to approximately \SI{6000}{J/mol} (corresponds to \SI{62}{meV/atom}).

Secondary GB dislocations generate a stress field that is nonlinear and contains non-elastic contributions within the dislocation core region, and is linear and elastic outside of the core. 
The dislocation core, which normally spans the length of two Burgers vectors \cite{Lazar2017}, serves as an essential trap for solutes. 
Notably, the core size is smaller than the binning scale used for our segregation energy quantification, which is approximately \SI{2}{nm} (see the line profile in \Fig{fig:4}{A \& B}).
Our main goal in this paragraph is to understand the gradient in these composition profiles. 
Thus, our analysis only focuses on the linear and elastic part of the stress field associated with the secondary GB dislocations.
The significant impact of secondary GB dislocations on solute segregation necessitates the introduction of an additional term to accurately represent the change in segregation energy.
The total segregation energy at the GB is expressed as:

\begin{equation}
\Delta G_{\text{seg}} = \Delta G^{\text{in}}_{\text{seg}} - P^{\text{XS}} \Delta V - T \Delta S^{\text{XS}}_{\text{seg}}.
\end{equation}
Here, $\Delta G^{\text{in}}_{\text{seg}}$ represents the intrinsic segregation energy of the solute at the GB, excluding contributions from secondary GB dislocations and accounting for the sum of the segregation enthalpy, elastic energy, and segregation entropy \cite{Han2016}, $P^{\text{XS}}$ denotes the hydrostatic pressure caused by the presence of secondary GB dislocations, $\Delta V$ represents the change in volume required for solute atoms to replace solvent atoms, and $\Delta S^{\text{XS}}_{\text{seg}}$ refers to the change in excess segregation entropy due to the introduction of secondary GB dislocations.
$\Delta G^{\text{in}}_{\text{seg}}$ depends on the kinematic degrees of freedom for a given GB \cite{Creuze2000, Wagih2020b}, while $\Delta S^{\text{XS}}_{\text{seg}}$ reflects the change in the configurations due to the presence of secondary GB dislocations.
We neglect the $\Delta S^{\text{XS}}_{\text{seg}}$ term and primarily attribute the modulation of segregation energy within a GB to the elastic energy resulting from dislocations at the GBs (see \Fig{fig:2}{C \& G}).

For bulk dislocations, the surrounding stress field can induce solute segregation in the areas surrounding the dislocations, known as Cottrell atmospheres \cite{Cottrell1949,Sun2018,Blavette1999,Thompson2007a}. 
In the case of secondary GB dislocations, the same applies, leading to a similar solute segregation phenomenon.
Unlike Burgers vectors in the bulk lattice, the unit Burgers vectors at GBs \cite{Hirth1973} are known as the displacement shift complete (DSC) vectors, shown in \Tab{tab:DSC} for GB2.
The pronounced contrast in secondary GB dislocations, seen in \Fig{fig:2}{C \& G}, suggests the formation of GB ledges through the agglomeration of individual secondary GB dislocations, as documented in previous studies \cite{Murr1970, Murr1975}.

We chose GB2 between grain $\alpha_4$ and grain $\alpha_5$ ($\Sigma 13b$) to calculate the elastic field around secondary GB dislocations (see \fig{fig:SFDSCa}{} and \fig{fig:SFDSCb}{}) using the Stroh formalism for anisotropic elasticity theory \cite{Ting1996, Stroh1958, Vattre2017} (see Methods \nameref{Simulation}).
The Stroh formalism requires a cutoff distance near the dislocation core, beyond which linear anisotropic elasticity theory applies and within which it fails. 
Here, a cutoff of \SI{0.5}{nm} was used for the stress field simulations. 
It is worth mentioning that non-linear anisotropic elasticity calculations by Lazar et al. \cite{Lazar2017} and atomistic simulations by Clouet et al. \cite{Clouet2009} are theoretical methods for estimating the stress field in the dislocation core.
Inputs for the stress field calculations were obtained directly from our 4DSTEM tomography analysis, i.e., the five kinematic degrees of freedom for GB2 (see \fig{fig:GBPlane}{A \& B} and \Tab{tab:Mis}).
\Figure{fig:4}{C \& D} show a periodic pattern in the stress field around the secondary GB dislocations with the DSC-lattice vectors $\mathbf{b_{\alpha5}} = \frac{a}{13} [\bar14\bar3]$ and $\mathbf{b_{\alpha5}} = \frac{a}{13} [\bar3\bar14]$ function as the Burgers vectors \cite{karakostas1979grain}.
Here, $a$ is the lattice constant of BCC Fe.

The orientation of the Burgers vector significantly influences the elastic field around the GB cores by varying its intensity and spread (see \Fig{fig:4}{C \& D}).
With the formula, $P^{\text{XS}} \Delta V$, we further calculated the extra contribution to the segregation energy from the elastic field produced by secondary GB dislocations, as presented in \Fig{fig:4}{E} for the quantitative measurements along the GB plane. 
The elastic energy contributes a modulation factor ranging from 0.4 to 2.5 times relative to the prediction of the Langmuir-McLean isotherm, corresponding to attenuation or enhancement depending on whether it takes a value below or above unity, respectively.

Our experiments confirm that the formation of secondary GB dislocations impacts W segregation at GBs in BCC Fe, resulting in an up to 100\% increase in the composition of GB segregation compared to a defect-free GB segment. 
Such secondary GB dislocations form patterns with regular spacing of approximately \SI{22}{nm}, which are topologically necessary to accommodate the GB's local curvature.
We have successfully quantified the segregation energy spectra of these GBs using interfacial excess mapping from APT data. 
The deviation in segregation energy within these spectra can reach magnitudes of up to approximately \SI{6000}{J/mol}. 
The extra elastic energy caused by secondary GB dislocations introduces a modulation factor, for example, ranging from 0.4 to 2.5 for the $\Sigma 13b$ GB, altering the local composition relative to predictions based on the Langmuir-McLean isotherm. 
These findings underscore the critical role of secondary GB dislocations not just as a topological necessity to accommodate GB shapes but also as deep solute traps, massively modulating solute segregation by up to a factor of two.
This enhances our understanding of GB decoration, offering new opportunities for the design of advanced alloys.

% Your references go at the end of the main text, and before the
% figures.  For this document we've used BibTeX, the .bib file
% scibib.bib, and the .bst file Science.bst.  The package scicite.sty
% was included to format the reference numbers according to *Science*
% style.

%BibTeX users: After compilation, comment out the following two lines and paste in
% the generated .bbl file. 

% \bibliography{scibib}

% \bibliographystyle{Science}

% \printbibliography[title={References}]
% % \printbibliography[title={References and Notes}, segment=\therefsegment]
% \end{refsegment}

\subsection*{Acknowledgments}
X.Z., W.L., and E.R. acknowledge funding by the German Research Foundation (DFG) through the project HE 7225/11-1.
X.C. and B.G. gratefully acknowledge the Collaborative Research Centre/Transregio (CRC/TRR) 270 HoMMage-Z01 project funded by the DFG.
The authors acknowledge Prof. Dr. Christian Liebscher for the discussion on transmission electron microscopy results.

\clearpage
  
\subsection*{Supplementary materials}
Materials and Methods\\
Supplementary Text\\
Figs. S1 to S10\\
Tables S1 to S3\\
Videos S1 to S3\\
References \textit{(28-71)}

\clearpage

\setcounter{figure}{0} % Reset figure counter
\setcounter{table}{0} % Reset table counter
% For Science format
\renewcommand{\thetable}{S\arabic{table}} % Change table numbering to "S1, S2, etc."
\renewcommand{\thefigure}{S\arabic{figure}} % Change figure numbering to "S1, S2, etc."

% \begin{appendices}
% \begin{refsegment}

\subsection*{Materials and Methods}\label{Method}

\subsubsection*{Sample preparation}\label{Sample}

\begin{figure}[ht]
\centering
    \includegraphics[width=16cm]{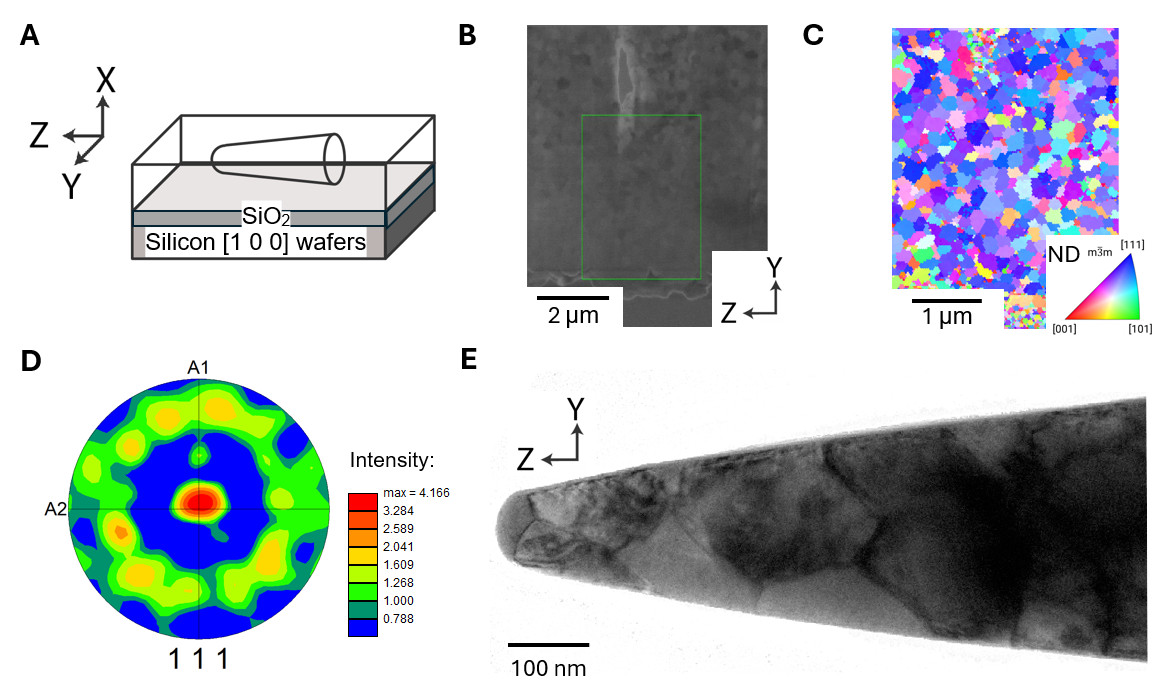}
    \caption{\textbf{Typical microstructure of the as-prepared Fe-1 at.\% W thin film.}
    (\textbf{A}) Definition of the coordination system used to describe the geometry of the Fe-1 at.\% W thin film specimen. 
    Here, the X-axis is aligned parallel to the thin film growth direction. 
    The Y and Z-axes, arbitrarily selected, lie in the plane perpendicular to the thin film growth direction. Once the needle-shaped specimen is prepared, the Z-axis serves as the tilting axis in TEM.
    (\textbf{B}) Region and (\textbf{C}) the corresponding orientation map for the transmission Kikuchi diffraction (TKD) analysis of the thin film prepared normal to the X-axis, termed as the plane view specimen. 
    (\textbf{D}) \hkl(111) pole figure, oriented parallel to the thin film growth direction, for the region shown in (C).
    (\textbf{E}) Bright field image of the as-prepared Fe-1 at.\% W APT sample.
   }
    \label{fig:MethodPVD}
\end{figure}

We deposited the model alloy Fe-1at.\%W thin film sample in a PVD cluster (BESTEC GmbH, Berlin, Germany).
The synthesis process involved co-sputtering a pure Fe target (\SI{99.995}{\percent}, Mateck, Germany) in a direct current cathode with a power of \SI{130}{W}, and a W target (\SI{99.95}{\percent}, Kurt J. Lesker, USA) in a radio frequency cathode at \SI{28}{W}, resulting in a total thickness of approximately \SI{2000}{nm}.
Prior to sputtering, the chamber was pumped to a base pressure of \SI{4.0e-8}{mbar}.
The Fe-W alloy thin films were then deposited at a pressure of \SI{5.0e-3}{mbar} and an Ar flux of \SI{40}{sccm} on smooth substrates of single crystalline silicon \hkl[100] wafers with a \SI{1.5}{\micro\meter} thermal $\text{SiO}_2$ diffusion and reaction barrier layer (see \fig{fig:MethodPVD}{A}), which was placed \SI{110}{mm} away from the sputtering targets.
The specimens were held at \SI{500}{\degree C} for \SI{240}{minutes} to facilitate solute segregation at GBs.

We prepared the needle-shaped specimen for correlative 4DSTEM tomography and APT characterization using the lift-out and annular milling technique developed by Thompson et al. \cite{Thompson2007b,DaCosta2024}.
The preparation of the specimen was conducted in a plasma-FIB (PFIB) instrument (FEI Helios PFIB) equipped with an Xe-ion source, which results in a low penetration depth that minimizes ion implantation and amorphization in the specimen.
The lift-out wedges were directly mounted onto an APT-compatible cylindrical Cu post, ready for $\pm$ \SI{90}{\degree} rotation along the tilting axis in the Fischione Model 2050 On-Axis Rotation Tomography Holder in the TEM.
During the lift-out procedure, we carefully aligned the rotation axis of the sample with that of the Cu post to match the tilt axis of the tomography holder.
A final PFIB milling condition of \SI{10}{pA} at \SI{5}{\keV} was applied to shape the needle-shaped specimen to a top diameter of \SI{100}{nm} (see \fig{fig:MethodPVD}{E}), while ensuring a clean surface.

\subsubsection*{4DSTEM tomography}\label{TEM}

We conducted the 4DSTEM tomography characterization using a JEM-2200FS (JEOL Ltd.) microscope, which was operated at an accelerating voltage of \SI{200}{kV} and equipped with a \SI{10}{\micro m} condenser (CL1) aperture and a \SI{50}{eV} energy filter.
This 4DSTEM tomography method relies on virtual dark-field images reconstructed from 4DSTEM datasets acquired at different tilts \cite{Harrison2022,Harrison2024}. 
A schematic of the 4DSTEM tomography characterization techniques is presented in \fig{fig:Method4DSTEMAPT}{A}. 
We collected 17 4DSTEM scans at tilts ranging from \(- \SI{80}{\degree}\) to \(+ \SI{80}{\degree}\) with a tilt step increment of \SI{10}{\degree}.
During the scan, a quasi-parallel electron probe, approximately \SI{2}{nm} in diameter, traverses a selected area of \SI{100}{\nano\meter} $\times$ \SI{300}{\nano\meter} in a line-by-line manner with a \SI{1}{nm} step, capturing nanobeam diffraction patterns at each probe position using a 4k $\times$ 4k CMOS detector (TemCam-XF416-TVIPS) with a pixel dwell time of \SI{41}{\milli\second}.

\begin{figure}[ht]
\centering
\includegraphics[width=16cm]{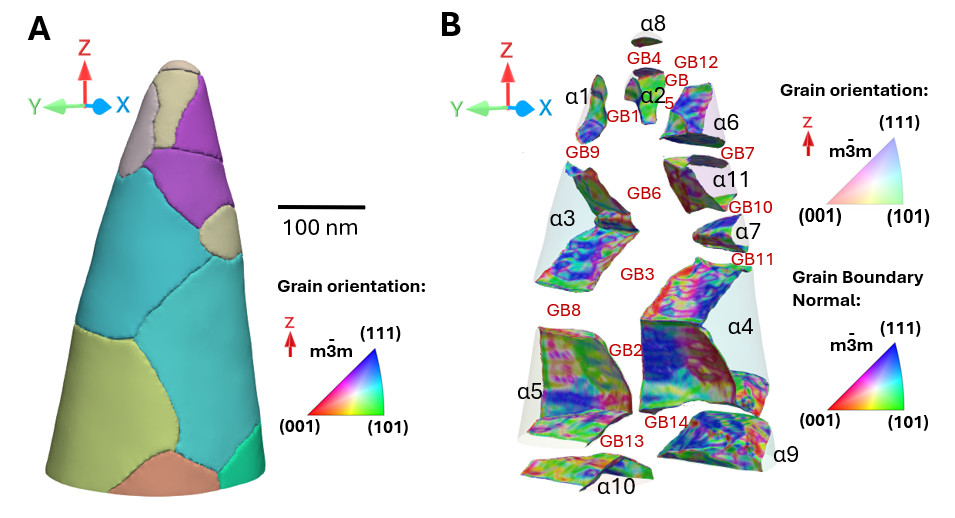}
    \caption{\textbf{3D crystallographic reconstruction of grains and GBs in the correlative Fe-1at.\%W needle-shaped specimen.} 
    (\textbf{A}) The orientations of grains and (\textbf{B}) the crystallographic information of GBs, both from the same sample as shown in \Fig{fig:1}{I}. 
    In (B), the local normal to the GB plane has been added, and grains have been manually detached to enhance clarity.
    Each GB features two plane normal mappings between adjacent grains.
    The coordinate system for each GB plane normal mapping is referenced to the respective grain involved in the mapping.
    }
    \label{fig:GBPlane}
\end{figure}

The phase and orientation of the grains in the 4DSTEM datasets were determined automatically using the multi-index algorithm, which retrieves information about overlapping grains via the ASTAR software package \cite{Rauch2010,Rauch2021} (see \fig{fig:GBPlane}{}).
This crystallographic information serves as a guideline for 1) precisely refining the tilting angles among different tilts; and 2) automatically generating virtual dark-field images for each grain in every 4DSTEM scan using the frozen template algorithm \cite{Rauch2021,Harrison2022,Harrison2024}.
We conducted the coarse alignment of the virtual dark-field images manually in Tomviz \cite{Levin2018}, followed by fine alignment using the PyCorrectedEmissionCT (corrct) package \cite{vigano2018}.
These well-aligned virtual dark-field images of each grain were used for 3D tomographic reconstruction through the Simultaneous Iterative Reconstruction Technique (SIRT) algorithm, which included 15 iterations combined with a non-negative minimum constraint to promote physical solutions.
For accurate determination of GB locations in volumes of overlapping grains, a max-pooling type approach is applied, which assigns the highest intensity volumes to the corresponding grains.
We then employed the marching cubes algorithm to generate normals for the planes corresponding to the two grains adjacent to each GB, thus providing distinct plane definitions for both \cite{Scikit2014}.
We used Blender \cite{Blender2018} to animate the method and visualize the reconstructed volume, as demonstrated in Videos s1-s3.

\subsubsection*{Correlative atom probe tomography}\label{APT}

\begin{figure}[ht]
\centering
    \includegraphics[width=16cm]{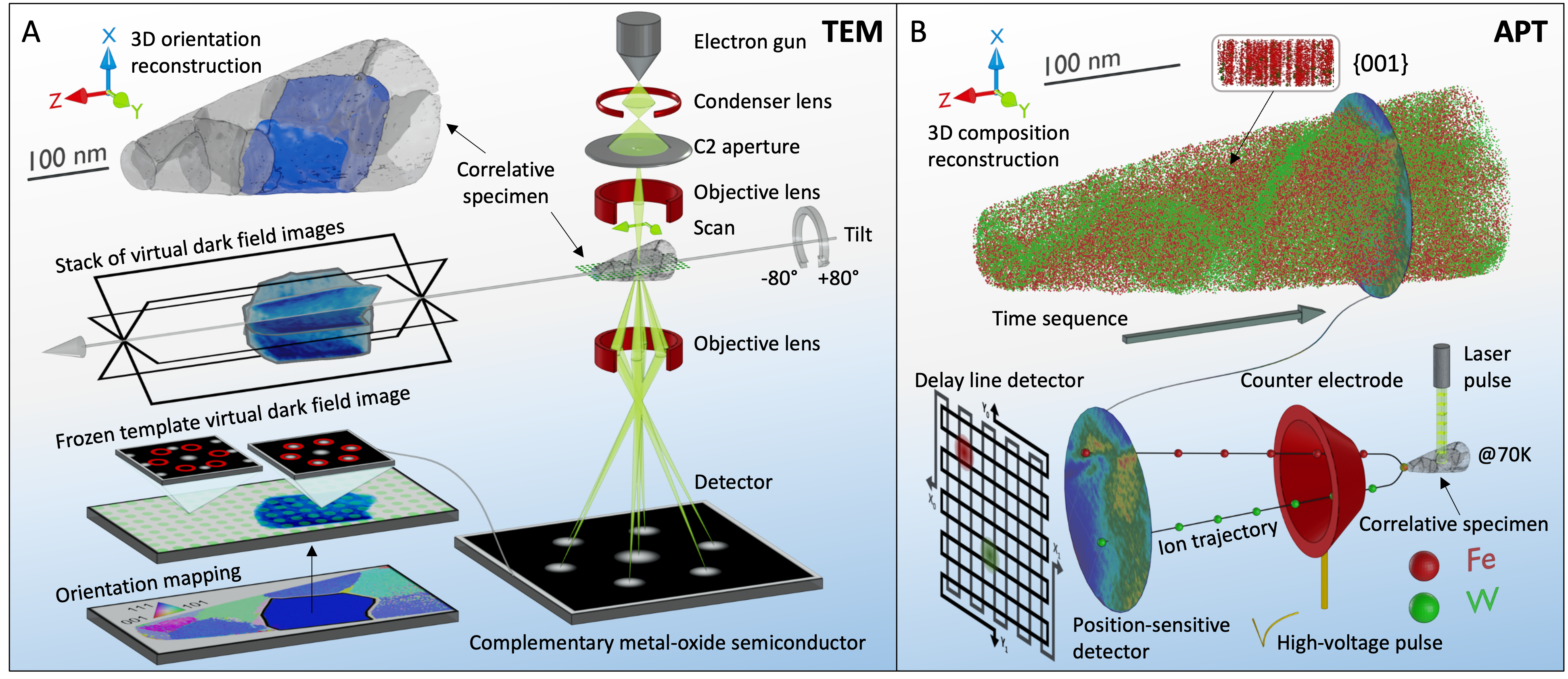}
    \caption{\textbf{Schematics of the correlative tomography characterization techniques}.
    (\textbf{A}) View of the internal optics, not to scale, inside a TEM column optimized for 4DSTEM data collection. 
    The process involves collecting multiple 4DSTEM datasets at various tilts to generate a series of virtual dark-field images for all grains using the frozen template algorithm \cite{Harrison2022,Harrison2024}. 
    Employing our in-house developed SPED3D software \cite{Harrison2022,Harrison2024}, we achieved 3D reconstructions, with the phases and orientations of all grains quantified.
    (\textbf{B}) View of the internal setup, not to scale, inside an APT analysis chamber configured with a straight flight-path. 
    Following TEM analysis, the specimen is transferred to the chamber and undergoes field evaporation ion by ion, with the coordinates and mass-to-charge ratio of each ion recorded for detailed 3D reconstruction and chemical analysis. 
    \(X_0, X_1, Y_0\) and \(Y_1\), times obtained from the delay-line detector used to determine the ion impact positions.
    }
    \label{fig:Method4DSTEMAPT}
\end{figure}

We analyzed the chemistry of the same needle-shaped specimen using APT after acquiring the 4DSTEM datasets.
Prior to loading the specimen into the APT analysis chamber, it was cleaned with low-kV Ar ions using a Gatan PECS Model 682 system at \SI{2}{\kilo\volt} and \SI{32}{\micro\ampere} to remove hydrocarbon layers accumulated during the TEM measurements.
We then conducted the APT measurements using a LEAP 5000XS instrument (Cameca Instruments) operated with a specimen temperature of \SI{70}{\kelvin} and a laser pulse energy of \SI{100}{\pico\joule} at a pulse repetition rate of \SI{200}{\kilo\hertz} for a \SI{2}{\percent} ions per pulse detection rate.
As shown in \Fig{fig:1}{J} and \fig{fig:Method4DSTEMAPT}{B}, we successfully collected 300 million ions from this correlative specimen, a critical volume that enables the simultaneous investigation of multiple GBs.
Data reconstruction was performed using the AP Suite 6.3 software package, following a calibration procedure to achieve the correct image compression factor and k-factor, which are essential for accurately shaping and spacing the lattice in the reconstructed volume \cite{Gault2011}.

\begin{figure}[ht]
\centering
\includegraphics[width=16cm]{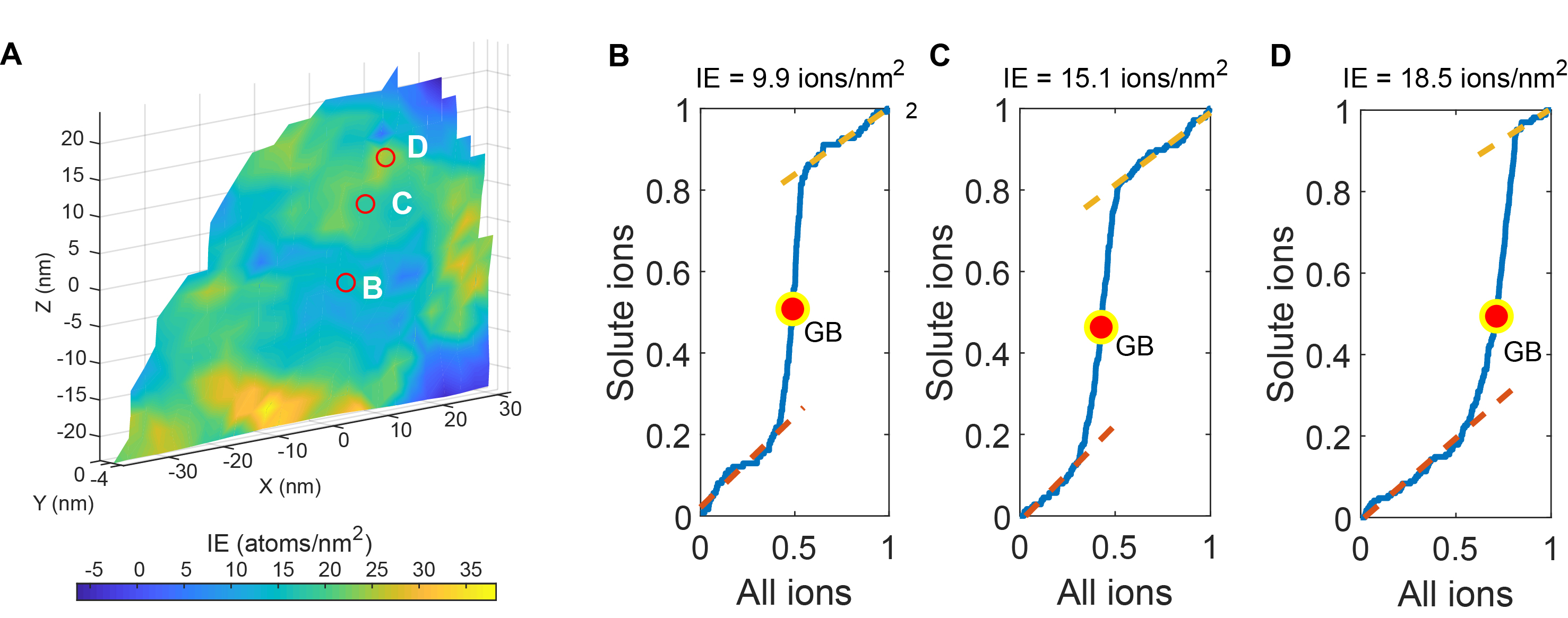}
    \caption{\textbf{Visualization of solute segregation by IE with integral profiles across the GBs}
    (\textbf{A}) IE mapping on GB1 with specific areas labeled B, C, and D.
    (\textbf{B-D}) Integral profiles across GB1 illustrate the segregation profiles of solutes for selected regions B through D. 
    These diagrams plot the fraction of solute ions against the total number of ions counted within a cylindrical-like volume across the interface, for areas B through D as shown in (A) \cite{Zhou2022}. 
    In these subfigures, blue lines represent experimental results, while orange and red dashed lines indicate theoretical fittings for adjacent two grains.
    Yellow circles mark the GB positions, highlighting regions of significant atom accumulation.
    }
    \label{fig:IE_ladder}
\end{figure}

We employed the APT\_GB software \cite{Zhou2022} to analyze the in-plane chemical distribution of W atoms at the GBs. 
With the pre-trained convolutional neural network, we automatically identified the locations of GB planes from the APT dataset, which were represented as triangular meshes with an average unit size of approximately \SI{7}{\nano\meter\squared}.
We generated integral profiles across the GBs to quantify the IE at each vertex on the mesh, with examples shown in \fig{fig:IE_ladder}{}). 
The resulting IE map is shown in \Fig{fig:3}{B}.

\subsubsection*{Displacement shift complete lattice of the $\Sigma13b$ grain boundary}\label{DSC}

\begin{figure}[ht]
\centering
\includegraphics[width=16cm]{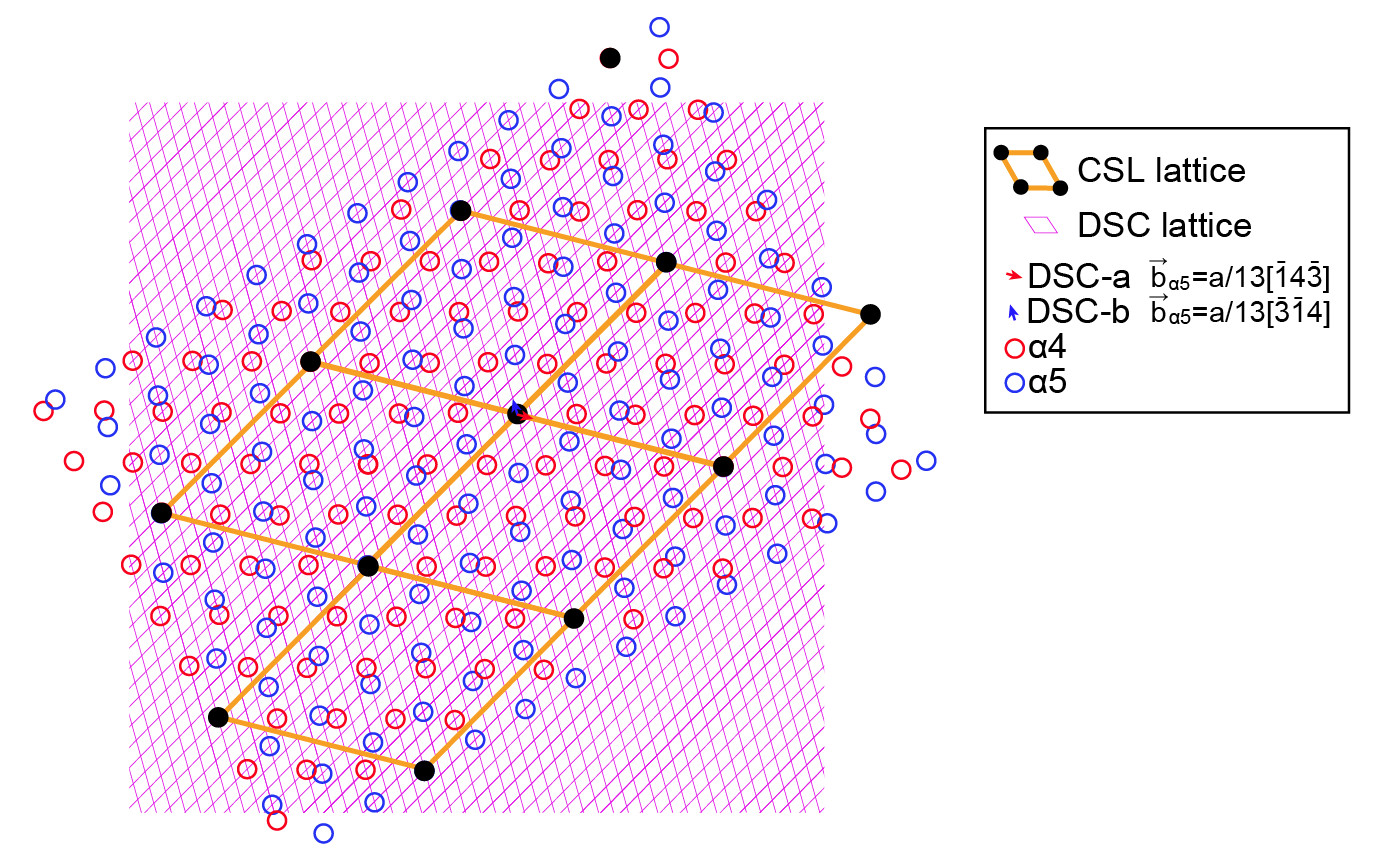}
    \caption{\textbf{The coincident site lattice (CSL) and DSC lattice of the $\Sigma13b$ GB}.
    The CSL lattice is represented by black dots, forming a foundational grid. 
    Overlaying this, the DSC lattice is depicted by pink lines.
    The red and blue circles signify atomic positions for $\alpha4$ and $\alpha5$, respectively.
    The red and blue arrows indicate the Burgers vectors of DSC-b, $\frac{a}{13}[\bar{3}\bar{1}4]$, and DSC-a, $\frac{a}{13}[\bar{1}4\bar{3}]$, respectively.}
    \label{fig:SFDSC}
\end{figure}

The dichromatic pattern of the DSC lattice of the $\Sigma13b$ GB (see \fig{fig:SFDSC}) and the analysis of Burgers vectors for secondary GB dislocations were obtained using the DSC lattice through O-lattice theory \cite{Bollmann1967}, utilizing an in-house developed open-source GB code \cite{Hadian2018}.

\subsubsection*{Stress field simulation}\label{Simulation}

Misfit dislocations on GBs can induce a stress field that would potentially change the segregation behavior of solute atoms in the alloy system. 
Such stress field was studied in previous work \cite{Abdolrahim2016,Vattre2017}, where a Stroh formalism is used to obtain the corresponding periodic stress field generated by semicoherent boundary and the required misfit dislocations. 
In essence, we solve the mechanical equilibrium equation written as 

\begin{equation}\label{eq:equil}
	\sigma_{ij,j} (x_{1}, x_{2}) = C_{ijkl} \, u_{k,jl} (x_{1}, x_{2}) = 0
\end{equation}
in both upper and lower grain with the critical interface condition expressed as
\begin{equation}\label{eq:bc}
	\begin{aligned}
		\relax[[\sigma_{2k} (x_{1}, x_{2} = 0 )]] & = 0 \\
		[[u_{k} (x_{1}, x_{2} = 0)]] & = - \sum_{n=1}^{\infty} \frac{1}{\pi n} \sin \left( \frac{2\pi n}{d} x_{1} \right) b_{k}
	\end{aligned}
\end{equation}
where $[[f]]$ denotes the jump function across the interface (interface normal is $x_{2}$ axis), $d$ is the dislocation spacing, $b_{k}$ is dislocation Burgers vector. 
This condition ensures that traction is in equilibrium across the interface and the relative displacement is consistent with our interfacial dislocation pattern. 

\begin{figure}[ht]
\centering
\includegraphics[width=16cm]{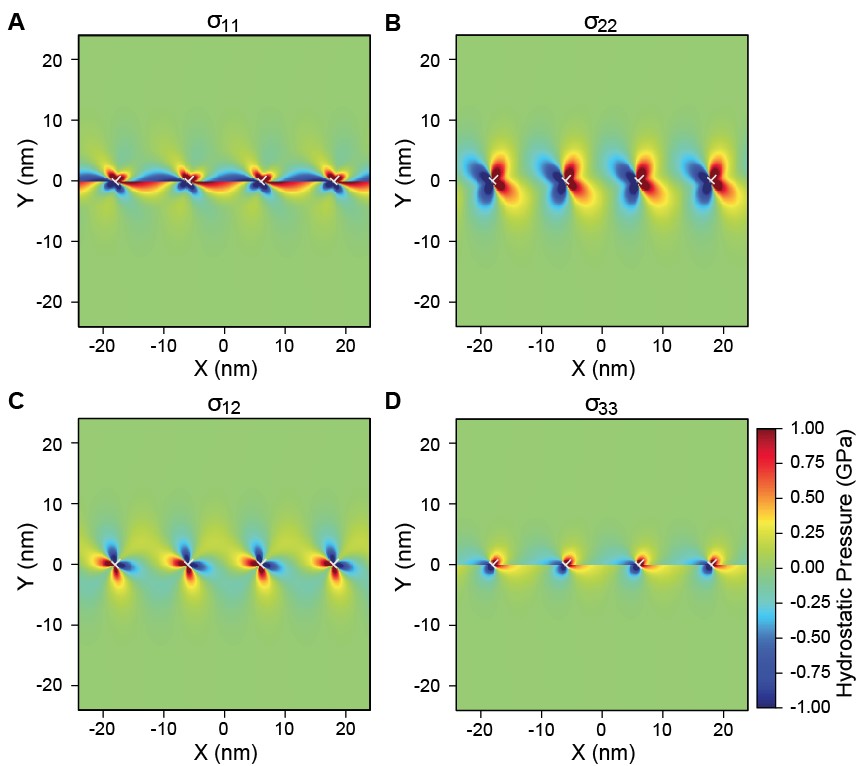}
    \caption{\textbf{Stress component distribution analysis for the $\Sigma13b$ GB with a DSC Burgers Vector of $\frac{a}{13}[\bar{1}4\bar{3}]$}:
    \textbf{(A)} $\sigma_{11}$, \textbf{(B)} $\sigma_{22}$, \textbf{(C)} $\sigma_{12}$, and \textbf{(D)} $\sigma_{33}$.}
    \label{fig:SFDSCa}
\end{figure}

\begin{figure}[ht]
\centering
\includegraphics[width=16cm]{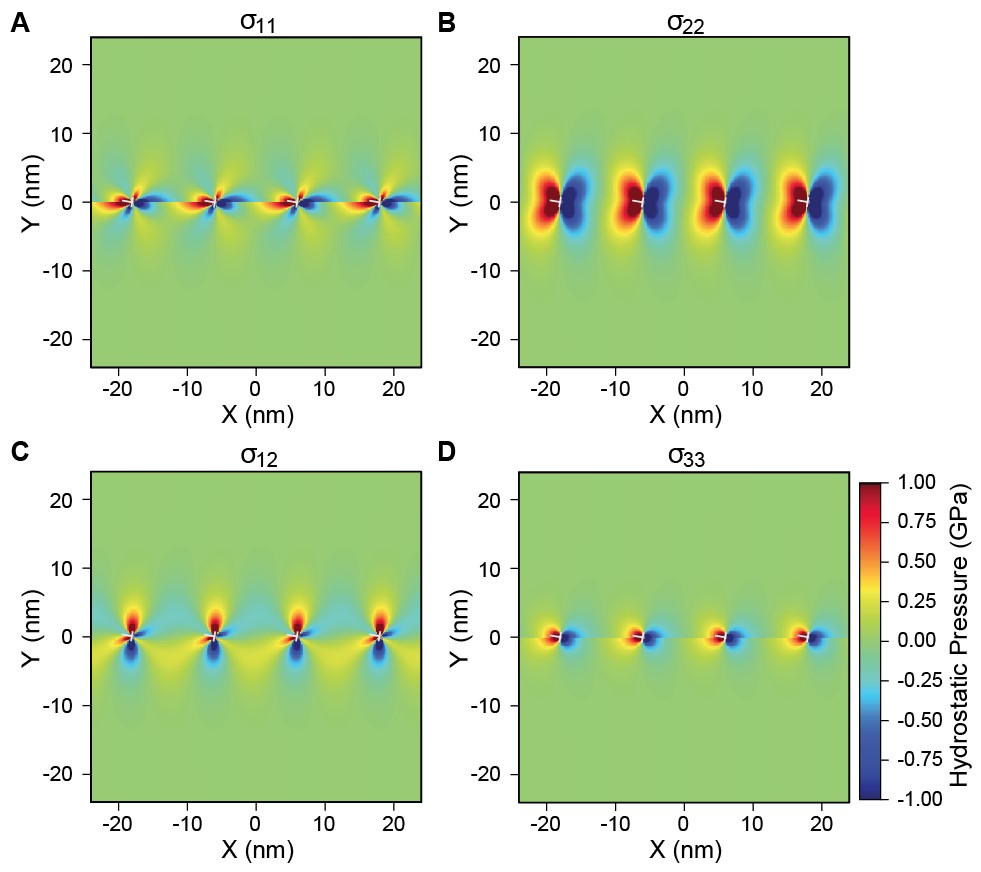}
    \caption{\textbf{Stress component distribution analysis for the $\Sigma13b$ GB with a DSC Burgers Vector of $\frac{a}{13}[\bar{3}\bar{1}4]$}:
    \textbf{(A)} $\sigma_{11}$, \textbf{(B)} $\sigma_{22}$, \textbf{(C)} $\sigma_{12}$, and \textbf{(D)} $\sigma_{33}$.}
    \label{fig:SFDSCb}
\end{figure}

To solve the equilibrium \eqref{eq:equil} on a periodic domain, Fourier series are used to write displacement field.
\begin{equation}
	u_{k} (x_{1}, x_{2}) = 2 \text{Re} \sum_{n=1}^{+\infty} e^{i \frac{2\pi n}{d} x_{1}} \tilde{u}_{k} (x_{2})
\end{equation}
where $\tilde{u}_{k} (x_{2})$ is the Fourier coefficient depending on position $x_{2}$. 
Substituting the displacement field into mechanical equilibrium together with further manipulation, the system boils down to a sextic equation which is the same with the one in a common Stroh formalism. 
To satisfy interfacial boundary condition, six complex scaling parameters can be assumed, which will then transform \eqref{eq:bc} to a set of six independent linear equation. 
Once the six complex scaling parameters are solved, the stress field can be obtained by calculating the strain with the solved displacement field and multiplying with the elastic constant in each grain (see \fig{fig:SFDSCa}{} and \fig{fig:SFDSCb}{}).

Area size: \( L = 48\times48\, \text{nm}^2 \);
Number of grid points: \( 500\times500 \);
Upper grain Euler angles: \( [41, 54.7, 45] \);
Lower grain Euler angles: \( [8, 54.7, 45] \);
Elastic constants of BCC Fe \cite{garrett2021}:
 \( C_{11} = 257.7 \times 10^9 \, \text{Pa} \),
\( C_{12} = 144.0 \times 10^9 \, \text{Pa} \),
\( C_{44} = 94.9 \times 10^9 \, \text{Pa} \); Poisson's ratio: \( \nu = 0.3 \);
Burgers vector of the dislocation: \( b = 0.083 \, \text{nm} \);
The core size of dislocations is defined as 2 Burgers vectors.

\clearpage

\subsection*{Supplementary Figures}\label{SI}

\clearpage

\clearpage

\clearpage

\begin{figure}
\centering
\includegraphics[width=12cm]{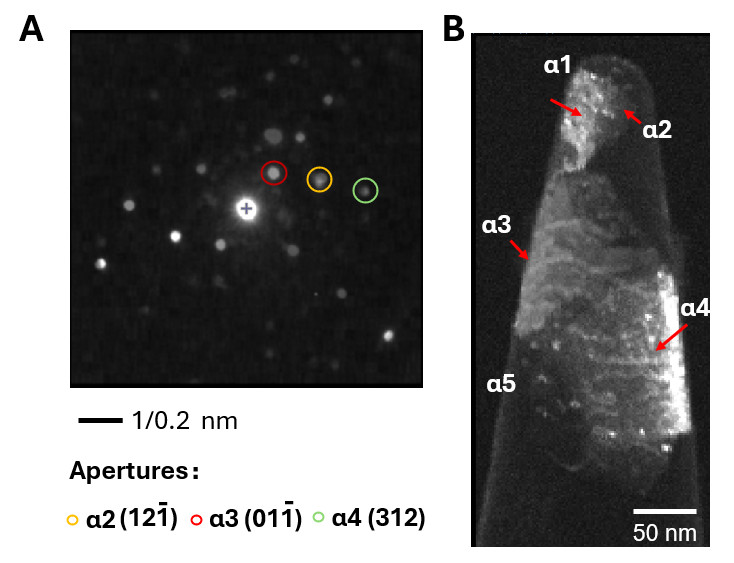}
    \caption{\textbf{Dislocation analysis for the Fe-1 at.\% W specimen.}
    (\textbf{A}) Displays the overlaid nanobeam diffraction patterns of grains $\alpha_2$ to $\alpha_4$. 
    (\textbf{B}) Shows a reconstructed 4DSTEM virtual dark-field image created using the virtual aperture depicted in (A), which includes potential diffraction spots calculated from grains $\alpha_2$ to $\alpha_4$.
    The 4DSTEM dataset captures nanobeam diffraction patterns at each scanning pixel, allowing for the use of a virtual aperture to select one or multiple spots from the recorded patterns to reconstruct virtual dark-field images \cite{Rauch2021}. 
    The aperture set shown in (A) includes the $\mathbf{g=[12\bar1]}$ spot of grain $\alpha_2$ (orange circle), the $\mathbf{g=[01\bar1]}$ spot of grain $\alpha_3$ (red circle), and the $\mathbf{g=[312]}$ spot of grain $\alpha_4$ (green circle), effectively highlighting grains $\alpha_2$, $\alpha_3$, and $\alpha_4$. 
    The diffraction pattern in (A) for grain $\alpha_3$ illustrates a close two-beam condition. 
    The corresponding virtual dark-field images for $(\mathbf{01\bar1})$ reveal the contrast of dislocations, with Burgers vectors satisfying $\mathbf{b} \cdot \mathbf{g} \neq 0$ \cite{Rauch2014b}. 
    We further analyze the line contrast on grain $\alpha_4$ in (B) using the $\mathbf{b} \cdot \mathbf{g} \neq 0$ criterion, confirming that the line contrasts are due to dislocations.
    }
    \label{fig:MVDF}
\end{figure}
\clearpage

\begin{figure}[ht]
\centering
\includegraphics[width=16cm]{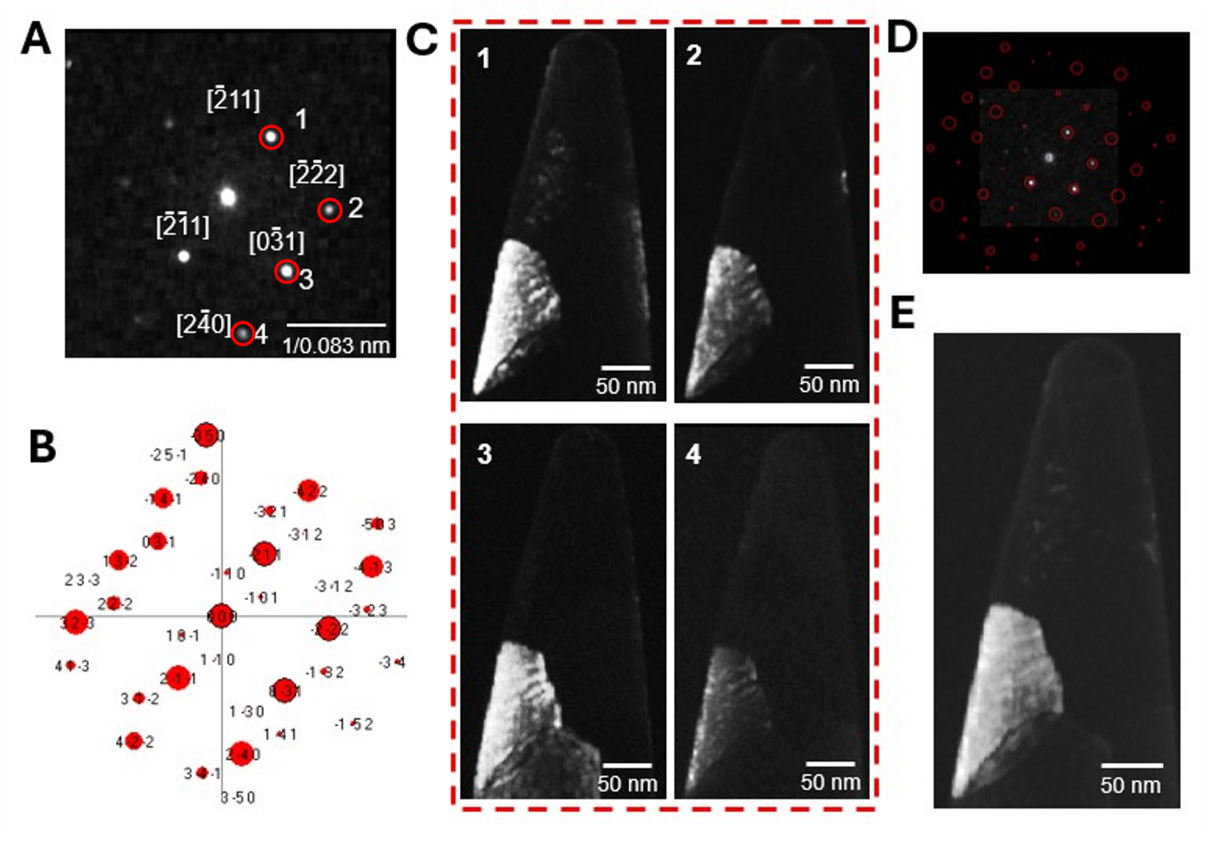}
    \caption{\textbf{Virtual dark-field reconstruction from the 4DSTEM dataset for grain $\alpha_5$.}
    (\textbf{A}) Diffraction pattern of grain $\alpha_5$. 
    (\textbf{B}) Indexing of the diffraction pattern shown in (A). 
    (\textbf{C}) Corresponding 4DSTEM dark-field images of the virtual aperture shown in (A). 
    (\textbf{D}) Template-aperture set based on the crystal orientation of grain $\alpha_5$. 
    (\textbf{E}) Template-aperture dark-field images. 
    }
    \label{fig:VDFG5}
\end{figure}
\clearpage

\begin{figure}[ht]
\centering
\includegraphics[width=16cm]{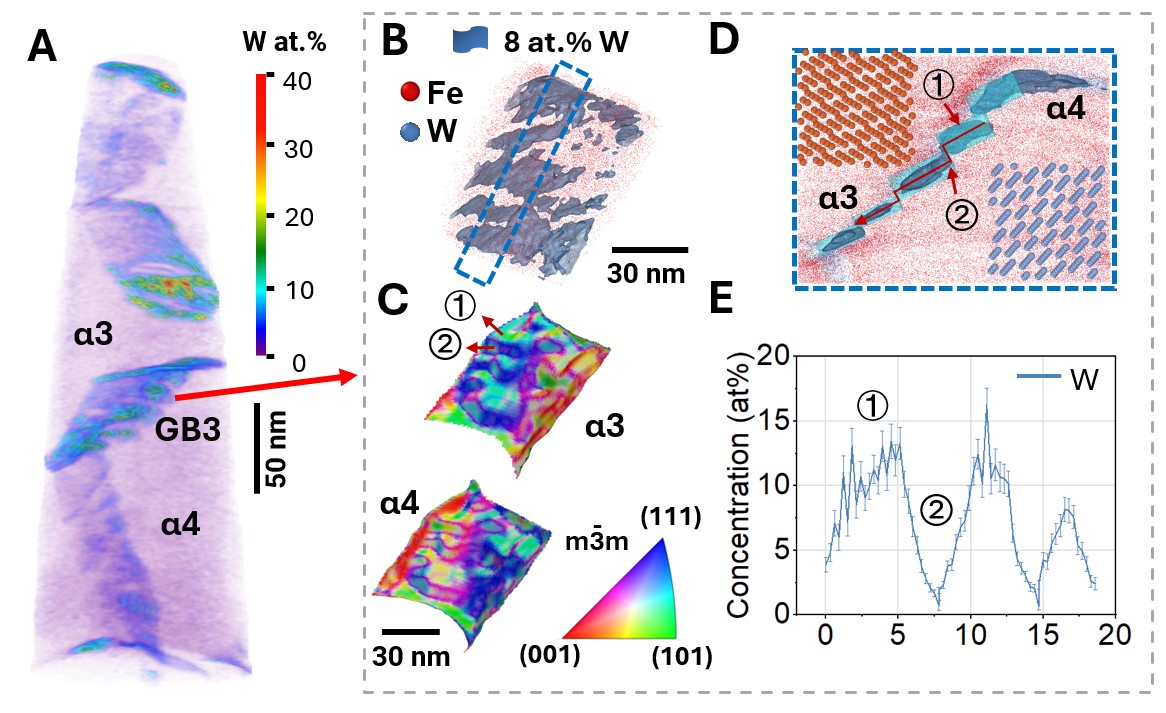}
    \caption{\textbf{Characterization of GB facets and their link to segregation patterns in the Fe-1 at.\% W specimen.}
    (\textbf{A}) The same mapping as in \Fig{fig:2}{A}.
    (\textbf{B}) The isosurfaces superimposed at 8.0 at.\% W on the atom maps of Fe and W highlight the GB indicated by the red arrow in (A).
    (\textbf{C}) Mapping of the local normal to the GB plane between adjacent grains, from the grain $\alpha_3$ side and the grain $\alpha_4$ side, respectively.
    (\textbf{D}) Side view of the GB within the blue dashed-line frame shown in (B), illustrating a staircase GB segregation pattern and indicating GB facets.
    The superimposed arrays of atoms represent the crystalline lattices for grains $\alpha_3$ and $\alpha_4$.
    (\textbf{E}) Compositional profile of W along the red arrow in (D).
    The numbers \textcircled{1} and \textcircled{2} in (C-E) indicate the same positions across different plots. 
    }
    \label{fig:GB3}
\end{figure}
\clearpage

\clearpage

\clearpage

\clearpage

\subsection*{Supplemental Tables}\label{ST}
\small

\begin{table}[ht]
\centering
    \caption{Bunge Euler Angle ($\phi_1$, $\Phi$, $\phi_2$) in degrees of all grains at the tilting angle of \SI{0}{\degree} in TEM.}
    \begin{tabular}{c|c|c|c}
    \hline
    \textbf{Grain} & $\phi_1$ & $\Phi$ & $\phi_2$ \\ \hline
    $\alpha_1$ & 244.0 & 31.1 & 43.0 \\ \hline
    $\alpha_2$ & 122.7 & 34.7 & 239.3 \\ \hline
    $\alpha_3$ & 257.6 & 46.9 & 40.1 \\ \hline
    $\alpha_4$ & 37.3 & 25.9 & 282.2 \\ \hline
    $\alpha_5$ & -26.4 & 15.4 & 324.2 \\ \hline
    $\alpha_6$ & 273.0 & 26.4 & 17.5 \\ \hline
    $\alpha_7$ & -37 & 16.6 & 4.2 \\ \hline
    $\alpha_8$ & 57.5 & 29.2 & 292.0 \\ \hline
    $\alpha_9$ & -24.9 & 27.5 & 336.1 \\ \hline
    $\alpha_{10}$ & -32.1 & 12.3 & 315.3 \\ \hline
    $\alpha_{11}$ & 73.9 & 22.1 & 255.1 \\ \hline
    \end{tabular}
    \label{tab:GrainEuler}
\end{table}
\clearpage

\begin{table}[ht]
\centering
    \caption{Unit Burgers vectors of secondary GB dislocations computed for principal coincidence system of body-centered cubic (BCC) crystals}
    \begin{tabular}{c|c|c|c}
        \hline
        \(\Sigma\) values & \(\mathbf{b_1}\) & \(\mathbf{b_2}\) & \(\mathbf{b_3}\) \\
        \hline
        5 & \(\frac{a}{5} [012]\) & \(\frac{a}{5} [02\bar1]\) & \(\frac{a}{10} [531]\) \\
        \hline
        13b & \(\frac{a}{13} [13\bar4]\) & \(\frac{a}{13} [3\bar41]\) & \(\frac{a}{26} [391]\) \\
        \hline
        45b & \(\frac{a}{45} [5\bar4\bar2]\) & \(\frac{a}{5} [0\bar12]\) & \(\frac{a}{30} [5\bar17]\) \\
        \hline
    \end{tabular}
    \label{tab:DSC}
\end{table}
\clearpage

\begin{table}[ht]
\centering
    \caption{Misorientation axis and angle for selected CSL GBs.}
    \begin{tabular}{c|c|c|c}
        \hline
        GB & Misorientation Axis & Misorientation Angle $\theta$ (°) & Closest CSL \\
        \hline
        $\alpha_1|\alpha_2$ & [20 0 -3] & 34.9 & $\Sigma5$ [Dev:5.6] \\
        \hline
        $\alpha_3|\alpha_4$ & [21 10 -19] & 37.1 & $\Sigma45b$ [Dev:1.4] \\
        \hline
        $\alpha_4|\alpha_5$ & [23 -34 -32] & 29.7 & $\Sigma13b$ [Dev:3.4] \\
        \hline
    \end{tabular}
    \label{tab:Mis}
\end{table}
\clearpage
% \noindent {\bf Fig. S1.} Please do not use figure environments to set
% up your figures in the final (post-peer-review) draft, do not include graphics in your
% source code, and do not cite figures in the text using \LaTeX\
% \verb+\ref+ commands.  Instead, simply refer to the figure numbers in
% the text per {\it Science\/} style, and include the list of captions at
% the end of the document, coded as ordinary paragraphs as shown in the
% \texttt{scifile.tex} template file.  Your actual figure files should
% be submitted separately.

% \printbibliography[title={References}, segment=\therefsegment, resetnumbers=false]
% \end{refsegment}

% \bibliographystyle{plain}  % You can change 'plain' to another style like 'unsrt' or 'alpha' if needed
% \bibliography{scifile}  % This should match the name of your .bib file without the extension

\end{document}